\documentclass[preprint2]{aastex6}

\usepackage{natbib}
\usepackage{graphicx}
\usepackage{bm}
\usepackage{amsmath}
\usepackage{amsfonts}
\usepackage{amssymb}

\def\gtwid{\mathrel{\raise.3ex\hbox{$>$\kern-.75em\lower1ex\hbox{$\sim$}}}}
\def\ltwid{\mathrel{\raise.3ex\hbox{$<$\kern-.75em\lower1ex\hbox{$\sim$}}}}
\def\\{\hfil\break}

\def\sun{\hbox{$\odot$}}
\def\earth{\hbox{$\oplus$}}
\def\lesssim{\mathrel{\hbox{\rlap{\hbox{\lower4pt\hbox{$\sim$}}}\hbox{$<$}}}}
\def\gtrsim{\mathrel{\hbox{\rlap{\hbox{\lower4pt\hbox{$\sim$}}}\hbox{$>$}}}}

\def\arcdeg{\hbox{$^\circ$}}

\newcommand{\mamo}[1]{\mbox{$#1$}}
\newcommand{\unit}[1]{\ifmmode \:\mbox{\rm #1}\else \mbox{#1}\fi}
\newcommand{\mone}{\mamo{^{-1}}}

\newcommand{\kms}{\unit{km~s\mone}}

\newcommand{\kpc}{\unit{kpc}}

\renewcommand{\lg}{\mamo{\sb{LG}}}

\newcommand{\msun}{\mamo{M_{\sun}}}

\begin{document}

\title{Bayesian Mass Estimates of the Milky Way: \\
including measurement uncertainties with hierarchical Bayes}
\author{Gwendolyn M. Eadie\altaffilmark{1}}
\author{Aaron Springford\altaffilmark{2}}
\author{William E. Harris\altaffilmark{1}}

\altaffiltext{1}{Dept.\ of Physics \& Astronomy, McMaster University, Hamilton, ON L8S 4M1, Canada.}
\altaffiltext{2}{Dept.\ of Mathematics \& Statistics, Queen's University, Kingston, ON K7L 3N6, Canada.}

\email{eadiegm@mcmaster.ca}
\shorttitle{}
\shortauthors{Eadie, Springford, and Harris}

\begin{abstract}

We present a hierarchical Bayesian method for estimating the total mass and mass profile of the Milky Way Galaxy.  The new hierarchical Bayesian approach further improves the framework presented by \cite{2015EHW, eadie2016} and builds upon the preliminary reports by \cite{eadieJSM, eadieIAU}. The method uses a distribution function $f(\mathcal{E},L)$ to model the galaxy and kinematic data from satellite objects such as globular clusters (GCs) to trace the Galaxy's gravitational potential. A major advantage of the method is that it not only includes complete and incomplete data simultaneously in the analysis, but also incorporates measurement uncertainties in a coherent and meaningful way. We first test the hierarchical Bayesian framework, which includes measurement uncertainties, using the same data and power-law model assumed in \cite{eadie2016}, and find the results are similar but more strongly constrained. Next, we take advantage of the new statistical framework and incorporate all possible GC data, finding a cumulative mass profile with Bayesian credible regions. This profile implies a mass within $125$kpc of $4.8\times10^{11}\msun$ with a 95\% Bayesian credible region of $(4.0-5.8)\times10^{11}\msun$. Our results also provide estimates of the true specific energies of all the GCs. By comparing these estimated energies to the measured energies of GCs with complete velocity measurements, we observe that (the few) remote tracers with complete measurements may play a large role in determining a total mass estimate of the Galaxy. Thus, our study stresses the need for more remote tracers with complete velocity measurements.

\end{abstract}


\maketitle

\label{firstpage}

\section{Introduction}

In our two previous papers, \citet[][hereafter Paper I]{2015EHW} and \citet[][hereafter Paper II]{eadie2016}, we estimated the Galaxy's mass and mass profile using a new Bayesian method and the kinematic data of Milky Way globular clusters (GCs) and dwarf galaxies (DGs). Paper I laid the groundwork: we tested the method on simulated data and then applied the method to Milky Way satellite data in a preliminary analysis. A main advantage of the new Bayesian method ws that both complete and incomplete velocity vectors were included in the analysis simultaneously. Furthermore, the tests on simulated data showed that our Galactic mass estimates were insensitive to incorrect velocity anisotropy assumptions. Paper I incorporated an analytic Hernquist model (for simplicity and testing of the method), and used GCs and DGs as tracers of the Milky Way's potential. The satellites were assumed to follow the same spatial distribution as the dark matter. Despite the simplicity of the model, the results were in agreement with many other studies
\citep[see][for a comparison figure]{wang2015}.

The promising results of Paper I led us to implement an arguably more realistic model for the Milky Way in Paper II, in which the distributions of the dark matter and the Galactic tracers are allowed to differ. The Paper II model uses power-law profiles with different parameters for the dark matter and tracers, and also includes velocity anisotropy as a parameter. This model is explained in detail by \cite{evans1997}, and previous applications of the model to the Milky Way and other galaxies were completed by \citet[][note that the notations vary between Evans' and Deason's papers]{deason2011, deason2012,Deason2012ApJ}. 

Because the model includes a spatial profile for only a single population of tracers, we used GC kinematic data alone instead of a mixture of DGs and GCs. The results in Paper II suggested a mass estimate for the Milky Way that was significantly lower than the mass found in Paper I under the Hernquist model, but closer in agreement to recent studies which suggest a ``light'' Milky Way \citep[e.g.][]{gibbons2014}.

An issue that is not fully addressed in Paper I or II is the inclusion of measurement uncertainty. Measurement uncertainties can differ substantially from object to object, with some tracers having very precise radial velocities or proper motions and others having very imprecise ones.

Using a sensitivity analysis, we found in Paper I that measurement uncertainties can play a significant role in the mass estimate of the Galaxy, contributing up to $50\%$ of the uncertainty in the estimate. In addition, we found that certain individual objects had very high leverage. For example, when the single GC Palomar 3 was removed from the analysis, the mass estimate of the Galaxy decreased by more than 12\%. Thus, it seems prudent to include measurement uncertainties in a rigorous and consistent way when estimating the mass and mass profile of the Galaxy.

Here, we substantially improve upon Paper II by introducing a \emph{hierarchical} Bayesian method that includes the measurement uncertainties of proper motions and line-of-sight velocities in a measurement model. Preliminary tests of this method have been reported by \citet*{eadieIAU} and \citet*{eadieJSM} using the Hernquist model and data from GCs and DGs in Paper I, but here we apply the arguably more realistic tracer model from Paper II, and also use all of the available GC data.

\section{Method}\label{sec:methods}

In Papers I and II, we defined the posterior distribution from Bayes' theorem as $p(\bm{\theta}|\bm{y})$, where $\bm{\theta}$ is the vector of model parameters, and $\bm{y}$ is the vector of data. In practice, the posterior distribution is difficult to calculate directly, and Markov Chain Monte Carlo (MCMC) methods are used to sample a distribution that is proportional to the posterior distribution. We write this distribution as
\begin{align}\label{eq:Bayes}
 p\left(\bm{\theta}|\bm{y}\right) &\propto \prod_i^n p\left(y_i|\bm{\theta}\right)p\left(\bm{\theta}\right) \\
 &= \prod_i^n p\left((r_i, v_{r,i}, v_{t,i})|\bm{\theta}\right) p\left(\bm{\theta}\right).
\end{align}
Above, $r_i$, $v_{r,i}$, and $v_{t,i}$ represent the Galactocentric distance, radial velocity, and tangential velocity of the $i^{th}$ tracer (GC). We assume that the GC positions and velocities are independent of one another, conditional on the value of $\bm{\theta}$.

In Paper II, we defined $p\left((r_i, v_{r,i}, v_{t,i})|\bm{\theta}\right)$ by the \emph{distribution function} (DF). The model for the dark matter halo's gravitational potential follows a power-law profile of $\Phi(r) = \Phi_o r^{-\gamma}$, and the spatial number density profile of the tracers follows $\rho(r) \propto r^{-\alpha}$. Using the Eddington formula as described in \citet{binney2008}, the DF is found to be
\begin{equation}\label{eq:DFLfinal}
	f(\mathcal{E},L) = \frac{ L^{-2\beta}\mathcal{E}^{ \frac{\beta(\gamma-2)}{\gamma} + \frac{\alpha}{\gamma}-\frac{3}{2}} } {\sqrt{ 8\pi^{3} 2^{-2\beta}} \Phi_o^{-\frac{2\beta}{\gamma} + \frac{\alpha}{\gamma}}} \frac{
	\Gamma\left( \frac{\alpha}{\gamma} - \frac{2\beta}{\gamma}+ 1\right)}
	{\Gamma\left( \frac{\beta(\gamma-2)}{\gamma} + \frac{\alpha}{\gamma} -\frac{1}{2}\right)}
\end{equation}
where $\mathcal{E} = -v^2/2 + \Phi(r)$, $L = r v_t$, and the model parameters are $\bm{\theta} = (\Phi_o, \gamma, \alpha, \beta)$ (beware of notational differences between \cite {evans1997} and \cite{deason2011, deason2012,Deason2012ApJ}). The parameter $\beta$ is the standard anisotropy parameter, where the limits $\beta=1$ and $\beta\rightarrow -\infty$ correspond to completely radial or completely tangential orbital distributions for the tracers \citep{binney2008}. 

The DF in Equation~\ref{eq:DFLfinal} assumes a spherical and non-rotating system, and also requires that the relative energy $\mathcal{E}$ is greater than zero (i.e. that tracers are bound to the Galaxy). Under this model, the mass profile of the dark matter halo is,
\begin{equation}\label{eq:Mr}
	M(r) = \frac{\gamma\Phi_o}{G}\left(\frac{r}{\text{kpc}}\right)^{1-\gamma}
	\end{equation}
\citep{Deason2012ApJ}, which has the physical limits of an isothermal sphere ($\gamma\rightarrow0$) and a central point mass ($\gamma\rightarrow1$).
 
Equation~\ref{eq:DFLfinal} is written in the Galactocentric reference frame--- the frame in which the geometry of the model is the most straightforward. The GC kinematic data and their uncertainties, on the other hand, are measured in the Heliocentric reference frame. Although the mathematical transformation of velocity and position vectors from a Heliocentric frame to a Galactocentric frame is relatively straightforward, transforming \emph{uncertainties} from one frame to the other requires complex error propagation which is non-linear, and that likely results in non-Gaussian errors. Therefore, we employ a different approach to incorporating the measurement uncertainties using a hierarchical Bayesian model.

\subsection{Hierarchical Bayesian Model}\label{sec:df}

In Paper II, all stochasticity in $\{r_i, v_{r,i}, v_{t,i} \}$ was due to Equation~\ref{eq:DFLfinal}, and none was due to measurement uncertainty. The measured values of  $\{r_i, v_{r,i}, v_{t,i} \}$ were assumed to be the true values, which we conditioned upon to obtain the posterior distribution for $\bm{\theta}$, the model parameters.

Now, we include a model for measurement uncertainty. The approach starts with a slight change in perspective: instead of treating the measurements of position $r$, line-of-sight velocity $v_{los}$, and proper motions in right ascension ($\mu_{\alpha}\cos{\delta}$) and declination ($\mu_{\delta}$) as the \emph{true} values, we treat these measurements as \emph{samples} drawn from a distribution which depends on the true (but unknown) values. That is, the true values are now included as parameters in the model. These parameters, the \emph{true} Galactocentric position and Heliocentric velocity components, are denoted in \textcolor{blue}{blue} with a $\textcolor{blue}{\dagger}$ symbol,
\begin{equation}\label{eq:datapars}
	\bm{\textcolor{blue}{\vartheta}} = \left( \textcolor{blue}{r^{\dagger}},\textcolor{blue}{v^{\dagger}_{los}}, \textcolor{blue} {\mu^{\dagger}_{\delta}}, \textcolor{blue}{\mu_{\alpha}\cos{\delta}^{\dagger}}\right),
\end{equation} 
For a given GC, the \emph{measurements} are denoted as
\begin{equation}\label{eq:data}
	\bm{y}=(r, v_{los}, \mu_{\delta}, \mu_{\alpha}\cos\delta),
\end{equation}
and the measurement \emph{uncertainties} are denoted in \textbf{\textcolor{red}{red}}:
\begin{equation}\label{eq:deltas}
\bm{\textcolor{red}{\Delta}} = (\textcolor{red}{\Delta r}, \textcolor{red}{\Delta v_{los}}, \textcolor{red}{\Delta\mu_{\delta}}, \textcolor{red}{\Delta\mu_{\alpha}\cos{\delta}}).
\end{equation}

We assume that the measurements are samples drawn from Gaussian (normal) distributions centered on $\bm{\textcolor{blue}{\vartheta}}$, and the measurement uncertainties $\bm{\textcolor{red}{\Delta}}$ are taken to be standard deviations. For example, the measurement of the line-of-sight velocity is drawn from a normal distribution centered on the true line-of-sight velocity, with a standard deviation equal to the measurement uncertainty. In statistical terms, this is akin to saying that $V_{los}$ is a \emph{random variable}, normally distributed with mean $\textcolor{blue}{v^{\dagger}_{los}}$ and variance $\textcolor{red}{{\Delta v_{los}}^2}$:
\begin{equation}
V_{los} \sim \mathcal{N}(\textcolor{blue}{v^{\dagger}_{los}},\textcolor{red}{{\Delta v_{los}}^2})
\end{equation}
(where $\mathcal{N}(\mu, \sigma^2)$ represents the Normal distribution). With this assumption, the probability of obtaining a \emph{measurement} $v_{los}$ is
\begin{equation}
p(V_{los}=v_{los}|\textcolor{blue}{v^{\dagger}_{los}},\textcolor{red}{\Delta v_{los}}) = \frac{1}{\sqrt{2\pi\textcolor{red}{{\Delta v_{los}}^2}}} e^{-\frac{(v_{los}-\textcolor{blue}{v^{\dagger}_{los}})^2}{2\textcolor{red}{{\Delta v_{los}}^2}}}.
\end{equation}

The same Gaussian assumption is made for the probabilities of the other measurements $p( \mu_{\delta} |\textcolor{blue}{\mu^{\dagger}_{\delta}},\textcolor{red}{\Delta \mu_{\delta}})$, $p( \mu_{\alpha} \cos{\delta} |\textcolor{blue}{\mu_{\alpha} \cos{\delta}^{\dagger}},\textcolor{red}{\Delta \mu_{\alpha} \cos{\delta}})$, and $p(r | \textcolor{blue}{r^{\dagger}}, \textcolor{red}{\Delta r})$. We assume that measurement errors are independent given the true values, so that the probability of measuring all components of a GC's kinematic quantities is simply the product of the probabilities defined above. Thus, the total likelihood is
\begin{multline}\label{eq:Likelihood}
\mathcal{L}(\bm{y}| \textcolor{red}{\mathbf{\Delta}},\textcolor{blue}{\mathbf{\vartheta}}) = p(r|\textcolor{blue}{r^{\dagger}},\textcolor{red}{\Delta r}) p(v_{los}|\textcolor{blue}{v^{\dagger}_{los}}, \textcolor{red}{\Delta v_{los}}) \times
\\
 p(\mu_{\delta}|\textcolor{blue}{\mu^{\dagger}_{\delta}}, \textcolor{red}{\Delta\mu_{\delta}})
p(\mu_{\alpha}\cos{\delta}|\textcolor{blue}{\mu_{\alpha}\cos{\delta}^{\dagger}},\textcolor{red}{\Delta\mu_{\alpha}\cos{\delta}})
\end{multline}
\citep{eadieJSM, eadieIAU}. This defines our measurement model. We acknowledge that the two components of the proper motion measurements are not actually independent. Their correlation could be incorporated using a multivariate normal, but because these correlations are not usually reported, we do not pursue it here.

Equipped with an expression for the likelihood (Eq.~\ref{eq:Likelihood}), we next define prior distributions on the parameters. The prior distributions on  $\bm{\textcolor{blue}{\vartheta}}$ link the measurement model to the tracer/galactic mass model. The parameters $\bm{\textcolor{blue}{\vartheta}}$ represent the true positions and velocities, and we assume that these parameters have a prior distribution defined by Equation \ref{eq:DFLfinal}, i.e. the DF. Thus, the DF is the prior distribution on $\bm{\textcolor{blue}{\vartheta}}$, and is denoted in shorthand as $p(h(\bm{\textcolor{blue}{\vartheta}})|\bm{\theta})$, where $h$ is the transformation from Heliocentric to Galactocentric coordinates (Section~\ref{sec:transform}).

Because the DF (the prior distribution on $\bm{\textcolor{blue}{\vartheta}}$) has its own parameters $\bm{\theta}$, then \emph{hyperpriors} $p(\bm{\theta})$ must also be specified. Thus, for a single GC or tracer, Bayes' rule is written as
\begin{align}\label{eq:bayeswords}
	p(\bm{\theta}|\bm{y_i}, \bm{\textcolor{red}{\Delta_i}}) & \propto \mathcal{L}(\bm{y_i}| \bm{\textcolor{red}{\Delta_i}},\bm{\textcolor{blue}{\vartheta_i}}) \times p(h(\bm{\textcolor{blue}{\vartheta_i}})|\bm{\theta}) \times p(\bm{\theta})\\
	& \propto \text{ Likelihood } \times \text{Prior} \times \text{Hyperprior}
\end{align}
Assuming that the GCs are conditionally independent, the posterior distribution is proportional to
\begin{equation}
	p(\bm{\theta}|\bm{y}, \textcolor{red}{\bm{\Delta}}) \propto \prod_{i=1}^{N} \mathcal{L}(\bm{y_i}|\textcolor{blue}{\bm{\vartheta_i}}, \textcolor{red}{\bm{\Delta_i}}) p(h(\textcolor{blue}{\bm{\vartheta_i}})|\bm{\theta}) p(\bm{\theta}) .
\end{equation}

The hierarchical Bayesian specification above provides a couple of improvements to Papers I and II (Equation\ref{eq:Bayes}). First and foremost, measurement uncertainties are included in the analysis in a meaningful way. Second, whereas before only 89 of 157 GCs could be included\footnote{Mainly due to the GCs' locations, see Papers I \& II}, we can now include 143 GCs. The remaining 14 GCs are objects for which no measurements of velocity are available (see Table 4 in Paper II).

\subsection{Defining Priors and Hyperpriors}\label{sec:priors}

Defining priors in the Bayesian paradigm is an opportunity for the researcher to state prior knowledge, gained from previous studies, and prior assumptions about model parameters. For this study, we use the same prior distributions for the model parameters $\Phi_o$, $\gamma$, and $\beta$ that were used previously: uniform distributions with bounds given and justified in Paper II. These lower and upper bounds for $\Phi_o$, $\gamma$, and $\beta$ are (1, 200), (0.3, 0.7), and (-0.5, 1.0) respectively.

\begin{figure}
 \centering
	\includegraphics[scale=0.4]{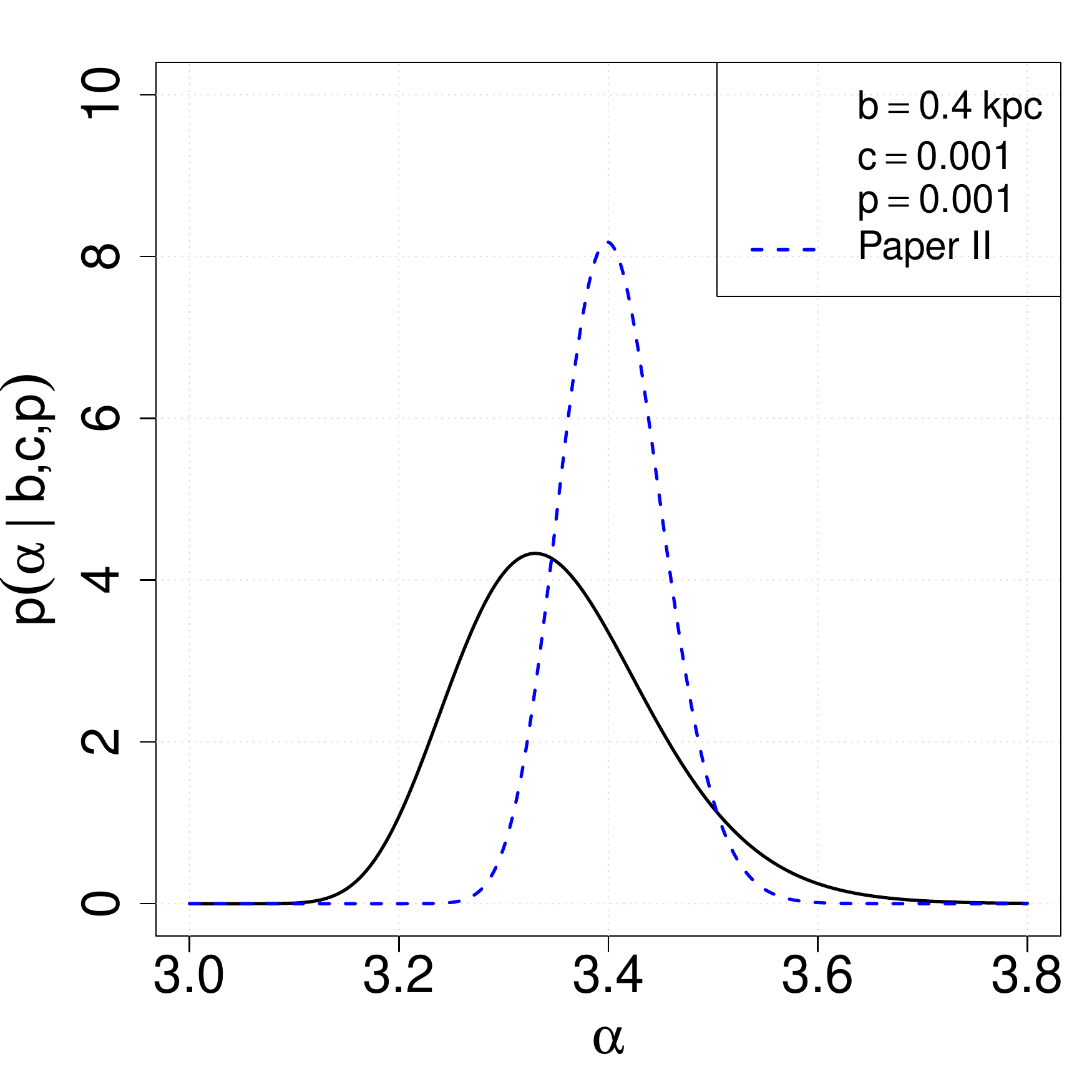}
	\caption{The solid black line is the prior distribution used in this paper, and the blue dashed line was the prior distribution used in Paper II. The solid-line prior probability distribution for $\alpha$ is determined using the extra GC data ($n=14$) that is not used in the rest of the analysis. Thus, the prior used in this study is less informative than that used in Paper II.}\label{fig:priora}
\end{figure}

The prior on the GC spatial distribution parameter, $p(\alpha)$, is a Gamma distribution. This choice was established and justified in Paper II, and was defined using the extra 68 GCs not otherwise included in the analysis. In this study, however, most of these previously excluded GCs can now be included because we do not have to depend on geometric assumptions to approximate $v_{los}$. There remain 14 GCs that are excluded in the data sample because they have only position measurements (see Section~\ref{sec:df}). We use these 14 GC positions to estimate and define a prior distribution for the parameter $\alpha$, in the same way as in Paper II. Figure~\ref{fig:priora} compares the new $p(\alpha)$ to that used in Paper II. Note that the new prior is wider than the one used in Paper II, because fewer GCs were used to estimate and define it. Including the extra GC data in the prior is akin to including the positions of all GCs in the analysis.

In summary, there are two sets of parameters in the hierarchical model: (1) the position and velocity parameters $\bm{\textcolor{blue}{\vartheta}$} and (2) the DF parameters $\bm{\theta}$. Bayes' theorem and the rules of conditional probability require a prior probability for both sets of parameters. The prior probability distribution for $\bm{\textcolor{blue}{\vartheta}}$ is Equation \ref{eq:DFLfinal} (the DF), because we assume that the positions and velocities come from the set of models determined by $\bm{\theta}$. The prior distributions on $\bm{\theta}$ are the \emph{hyperprior} distributions $p(\bm{\theta})$ described above and in Paper II.

\subsection{Transformation of Velocities}\label{sec:transform}

In this section we discuss the function $h(\bm{\textcolor{blue}{\vartheta}})$ first mentioned in Section~\ref{sec:df}. The $h(\bm{\textcolor{blue}{\vartheta}})$ notation symbolizes the transformation of velocity parameters in Eq.~\ref{eq:datapars} from a Heliocentric parameterization $(\textcolor{blue}{v^{\dagger}_{los}}, \textcolor{blue}{\mu^{\dagger}_{\delta}}, \textcolor{blue}{\mu_{\alpha}\cos{\delta}^{\dagger}})$ to a Galactocentric parameterization $(\textcolor{blue}{v^{\dagger}_r}, \textcolor{blue}{v^{\dagger}_{t}})$, following the method presented in \cite{johnson1987}. We review the \cite{johnson1987} method here for completeness and in order to highlight some important points.


The first step is to transform the Heliocentric velocities into Galactic space-velocities $(U,V,W)$ in a right-handed coordinate system: 
\begin{equation}
\left[\begin{array}{c}
	U\\
	V\\
	W
	\end{array}\right]	= \mathbf{T}\cdot\mathbf{A}\left[\begin{array}{c}
v_{los}\\
k\mu_{\alpha}cos(\delta) / \lambda \\
k\mu_{\delta} / \lambda
\end{array}\right]+\left[\begin{array}{c}
U_{\odot}\\
V_{\odot}\\
W_{\odot}
\end{array}\right]
\end{equation}
where $U$ is positive toward the Galactic center, $V$ is positive in the direction of Galactic rotation, and $W$ is positive above the Galactic plane. The solar motion is set to $(U_{\odot}, V_{\odot}, W_{\odot}) = (11.1, 12.24, 7.25)$ \citep{schonrich2010}, $k=4.74057$ (the equivalent in km/s of one AU in one tropical year), and $\lambda$ is the parallax (in arcsec) of the GC \citep{johnson1987}. The matrices $\mathbf{T}$ and  $\mathbf{A}$ depend on the right-ascension (R.A,) and declination (decl.) of the North Galactic Pole (as determined by the Hipparcos catalog) and GCs respectively, where
\begin{equation}\label{eq:T}
\mathbf{T} = \left[\begin{array}{ccc}
-0.0548755604 & -0.8734370902 & -0.4838350155 \\
+0.4941094279 & -0.4448296300 & +0.7469822445 \\
-0.8676661490 & -0.1980763734 & +0.4559837762
\end{array}\right]
\end{equation}
\citep{Hipparcos} and where $\mathbf{A}$ for a single GC is
\begin{equation}\label{eq:A}
\mathbf{A}=\left[\begin{array}{ccc}
+\cos\alpha\cos\delta & -\sin\alpha & -\cos\alpha\sin\delta\\
+\sin\alpha\cos\delta & +\cos\alpha & -\sin\alpha\sin\delta\\
+\sin\delta & 0 & +\cos\delta
\end{array}\right].
\end{equation}
Above, $\alpha$ and $\delta$ are the R.A. and decl. respectively, in decimal degrees (this $\alpha$ is of course different from the one used to parameterize the GC distribution above). We take the GCs' parallax, and R.A. and decl. positions as fixed, but treat the Galactocentric distance $\textcolor{blue}{r}$ as a parameter in the model, and assign an uncertainty of 5\% to the measured $r$ value \citep{1996harris,2010Harris}.

The next step is to transform the Cartesian, rotating Galactic frame velocity components $(U,V,W)$ into components in a cylindrical, non-rotating Galactocentric reference frame $\left(\Pi,\Theta,W\right)$. First the adjustment for the rotation of the Galaxy at $R_{\odot}=8.0$\kpc~ is taken to be 220km/s, to obtain $(U_{gc},V_{gc},W_{gc})$, and then this vector is transformed to a non-rotating, right-handed cylindrical system via

\begin{equation}\label{eq:PiThetaW}
\left[\begin{array}{c}
\Pi\\
\Theta\\
Z
\end{array}\right]=\left[\begin{array}{ccc}
\cos\theta & \sin\theta & 0\\
-\sin\theta & \cos\theta & 0\\
0 & 0 & 1
\end{array}\right]\left[\begin{array}{c}
U_{gc}\\
V_{gc}\\
W_{gc}
\end{array}\right].
\end{equation}

\begin{figure} 
\includegraphics[trim={0cm 0cm 1cm 1cm}, scale=0.69]{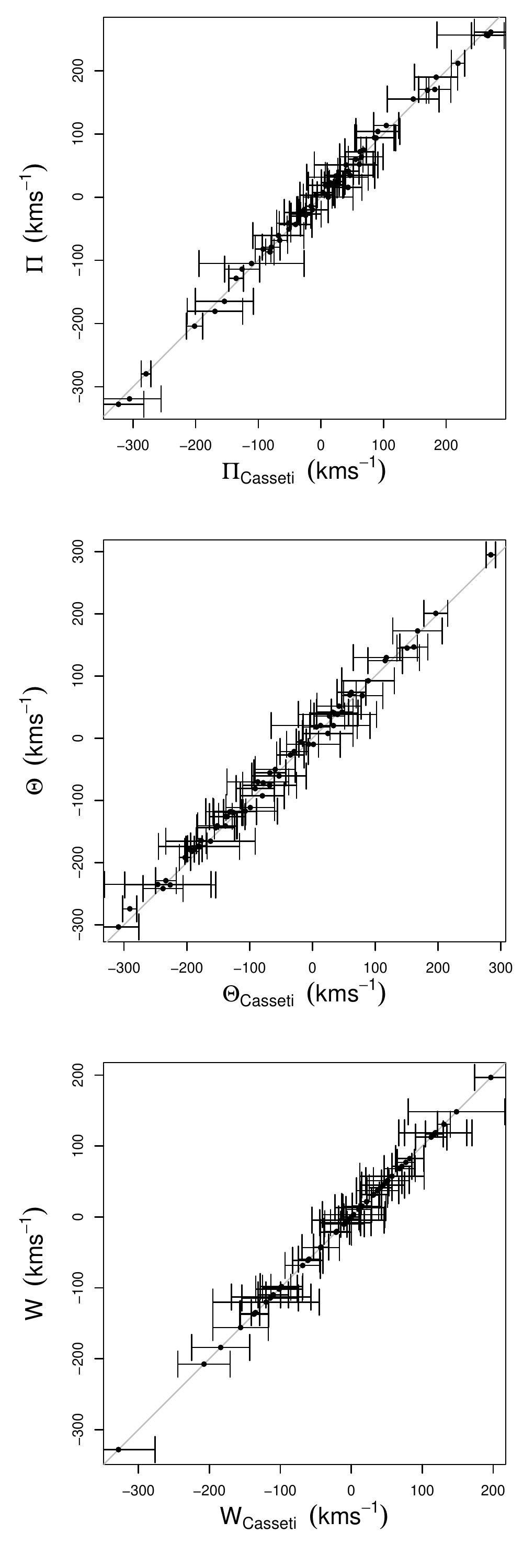}
	\caption{Our transformation from Heliocentric velocities to Galactocentric velocities ($\Theta, \Pi, W$) for GCs with proper motions, compared to the Casseti catalog. The abscissa are from the Casseti online catalog, and the ordinates are our transformation values. The error bars are the uncertainties given in the Casseti catalog, and the grey line has a slope of one.}
\label{fig:comparePTW}
\end{figure}

As a test of the entire transformation, we compare our derived $(\Pi, \Theta, W)$ to  the Casseti online catalog of GC velocity measurements (Figure~\ref{fig:comparePTW}) \citep{1999Dinescu, dinescu2004, dinescu2005, dinescu2010, dinescu2013}\footnote{Updated catalog: www.astro.yale.edu/dana/gc.html}.

Finally, the velocity components in Eq.~\ref{eq:PiThetaW} are transformed to the spherical coordinate system
\begin{equation}
\left[\begin{array}{c}
v_{r}\\
v_{\theta}\\
v_{\phi}
\end{array}\right]=\left[\begin{array}{ccc}
\cos\phi & 0 & \sin\phi\\
0 & 1 & 0\\
-\sin\phi & 0 & \cos\phi
\end{array}\right]\left[\begin{array}{c}
\Pi\\
\Theta\\
W
\end{array}\right]
\end{equation}
where $v_t^2=v_{\theta}^2 + v_{\phi}^2$. To reiterate, the complete transformation from the Galactocentric parameterization to the Heliocentric parameterization described above is represented by $h(\bm{\textcolor{blue}{\vartheta}})$ in Equation~\ref{eq:DFLfinal}.


%

\subsection{Improved Computational Methods}

The posterior distribution is sampled using the same MCMC method with a hybrid-Gibbs sampler that was utilized in Papers I and II. One improvement is that the proposal distributions for $\bm{\textcolor{blue}{\vartheta}}$ and $\bm{\theta}$ are determined using the \emph{adaptive} MCMC method described by \cite{2009robertsrosenthal}. A multivariate covariance matrix is determined for each GC's $\bm{\textcolor{blue}{\vartheta}}$ parameters, and for $\bm{\theta}$, via an iterative burn-in process. The advantage of the adaptive MCMC method is that the target posterior distribution is sampled much more efficiently by taking into account correlations between parameters. A second minor change is that we now run seven, independent parallel chains instead of three, thereby obtaining the same number of samples in less than half the time.

\section{Kinematic Data}\label{sec:data}

The kinematic data used in this study are presented in Table 1 of Paper II. In Paper II, only 89 out of 157 GCs were used in the analysis, mainly because the approximation $|v_{los}|\approx|v_r|$ did not hold for most GCs without proper motions. Other GCs were excluded in the analysis of Paper II, due to high reddening association with the Sagittarius dwarf galaxy, or no velocity measurements. 

As described in Sections~\ref{sec:methods} and \ref{sec:data}, using the hierarchical Bayesian framework allows all of the incomplete data to be included without having to make any geometric arguments like those used in Papers I and II, because the likelihood $\mathcal{L}$ is written in the Heliocentric frame. Now that we are accounting for uncertainties, we also include the GCs subject to high reddening. In the present analysis, the GCs associated with the Sagittarius dwarf do not change the result significantly and therefore we choose to include them. Altogether, this increases the size of the data set significantly, from 89 to 143 GCs.
\newpage
\section{Analysis Overview}

To make a fair comparison between the non-hierarchical method of Paper II and the hierarchical method presented here, and to thereby directly test the influence of measurement uncertainties, we first apply the hierarchical Bayesian method to the same kinematic data that was analyzed in Paper II (i.e. only 89 GCs). In this case, we use the prior distribution $p(\alpha)$ for the tracer spatial parameter that was used in Paper II (i.e. the dashed blue line in Figure~\ref{fig:priora}).

Next, we use the hierarchical Bayesian method with 143 GCs, using prior distribution $p(\alpha)$ defined by the extra 14 GCs without velocity measurements (solid line in Figure~\ref{fig:priora}).

\section{Results}\label{sec:results}
\begin{figure}[t]
\centering
	\includegraphics[totalheight=0.39\textheight]{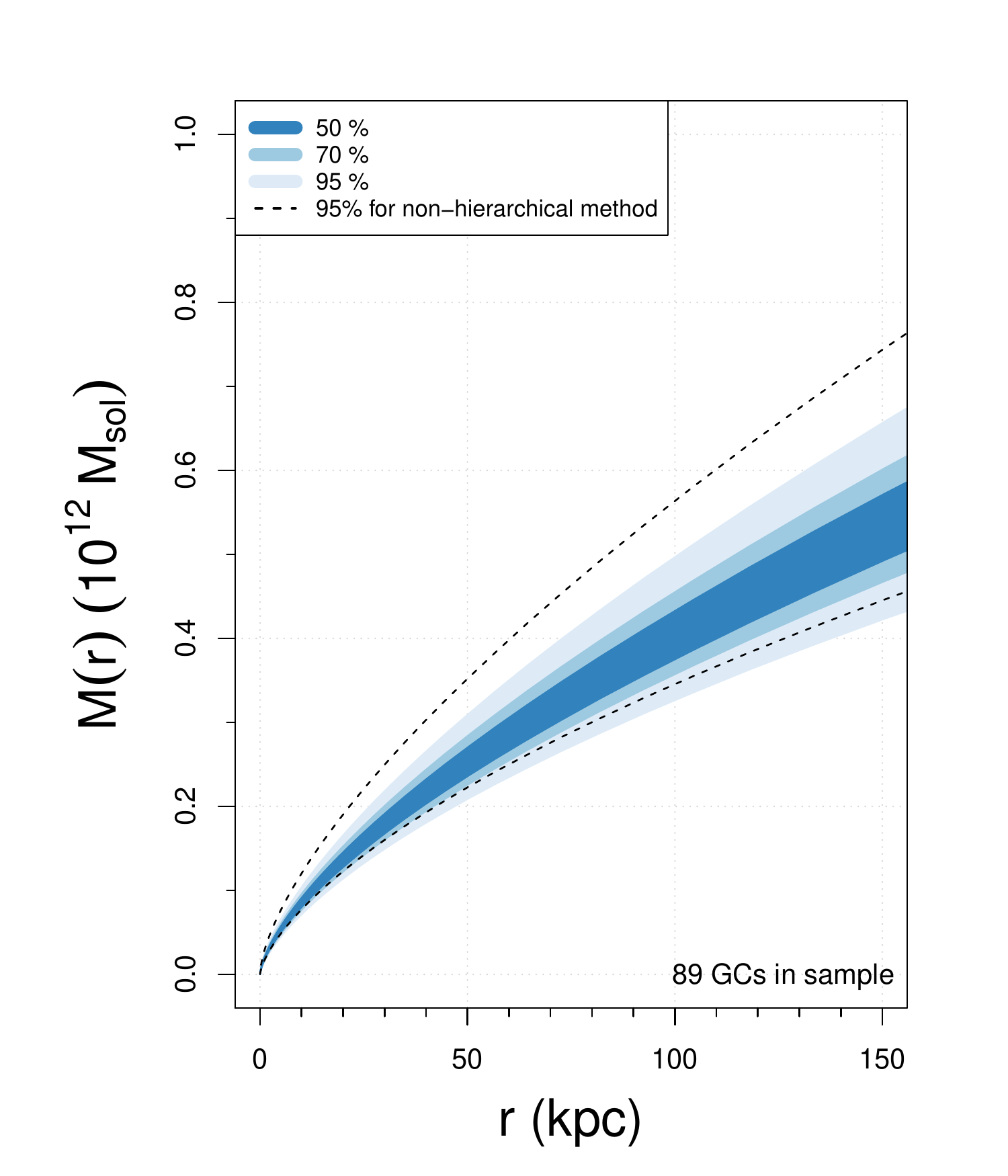}
	\caption{The blue shaded areas are the Bayesian credible regions for the cumulative mass profile of the Milky Way, using the hierarchical method and 89 GCs. The black dashed lines show the 95\% credible regions for the non-hierarchical method and 89 GCs (i.e. the results from Paper II).}
	\label{fig:89gcs}
\end{figure}

Figure~\ref{fig:89gcs} compares the 95\% Bayesian credible regions for the mass profiles of the Milky Way from Paper II (the dashed, black lines) to the 50\%, 75\%, and 95\% regions from the present paper (the shaded blue regions). Both results rely on the same 89 GC sample used in Paper II; the only difference between the two analyses is that measurement uncertainties are now included. The main result of including measurement uncertainties via the hierarchical method is a stronger constraint on the mass profile and mass estimate compared to the method used in Paper II.

\begin{figure}
\centering
	\includegraphics[totalheight=0.39\textheight]{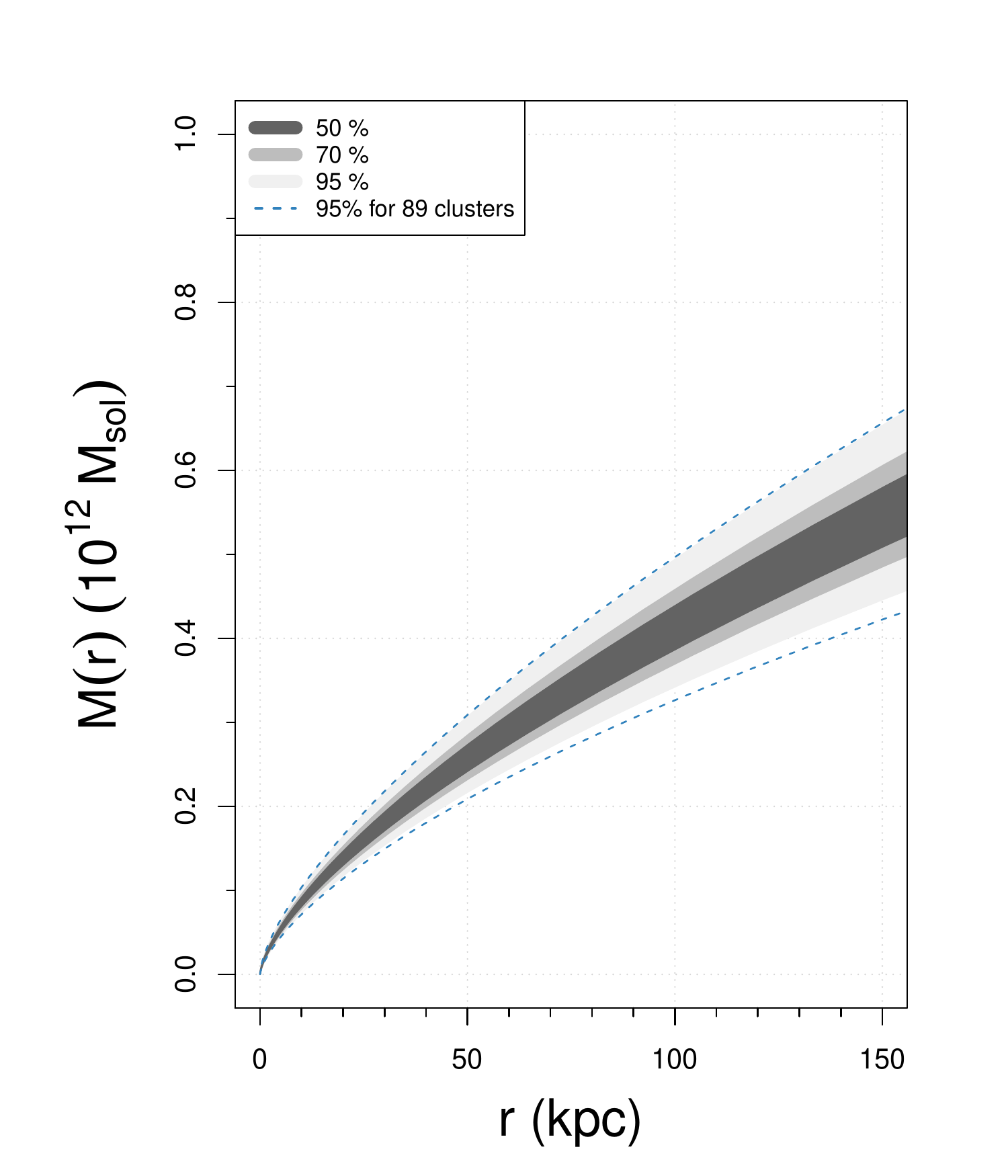}
	\caption{The grey shaded areas are the Bayesian credible regions for the cumulative mass profile of the Milky Way using the hierarchical method and 143 GCs. The blue dashed lines show the 95\% credible regions for the hierarchical method and 89 GCs (i.e. the outermost blue region in Figure~\ref{fig:89gcs}).}
	\label{fig:143gcs}
\end{figure}

As mentioned in Section~\ref{sec:df} and \ref{sec:data}, one advantage of the hierarchical model is that the GC sample size is increases from 89 to 143. Figure~\ref{fig:143gcs} compares the estimated mass profile using 89 GCs to the profile using 143 GCs. The dashed, blue lines indicate the 95\% Bayesian credible regions from Figure~\ref{fig:89gcs}, and the black shaded regions indicate the credible regions when 143 GCs are included.  The increase in total sample size likely accounts for the slight narrowing of the Bayesian credible regions. However, the difference between the hierarchical results from the 89 GC sample and those from the 143 GC sample is not as large as might be expected.


We conjecture that a more constrained estimate using the present method will require a higher proportion of complete data. Although the sample size increased by more than 60\%, the GC data that are added to the sample are all incomplete. Of the 89 GCs used in Paper II, 71 of these had complete velocity data. Thus, with 143 GCs the \emph{proportion} of GCs with complete velocity data decreased to about 50\%.

The GCs are subject to the total gravitational potential within their orbits, and thus trace the Galaxy's total mass out to 125kpc (the distance of the farthest GC in our sample). Using the hierarchical Bayesian method presented here, the power-law models, and the priors, and confronting this coherent model with data from 143 GCs returns a total mass within 125kpc of $4.8\times10^{11}\msun$, with a 95\% credible region of $(4.0, 5.8)\times10^{11}\msun$.


Extrapolating our mass profile in Figure~\ref{fig:143gcs} out to a virial radius that corresponds to 200 times the critical density of the universe, assuming $H_o=67.8\kms\text{Mpc}^{-1}$ \citep{Planck2015}, we find that $r_{200}=179~(164,194)\kpc$ and $M(r_{200})=6.2~(4.7, 7.8)\times10^{11}\msun$ (the numbers in brackets correspond to the 95\% Bayesian credible regions). Extrapolating out further, we find that the mass within $300\kpc$ is $M(300\kpc)=0.9~(0.7, 1.1)\times 10^{12}\msun$.

\begin{figure}[t]
\centering
	\includegraphics[totalheight=0.35\textheight]{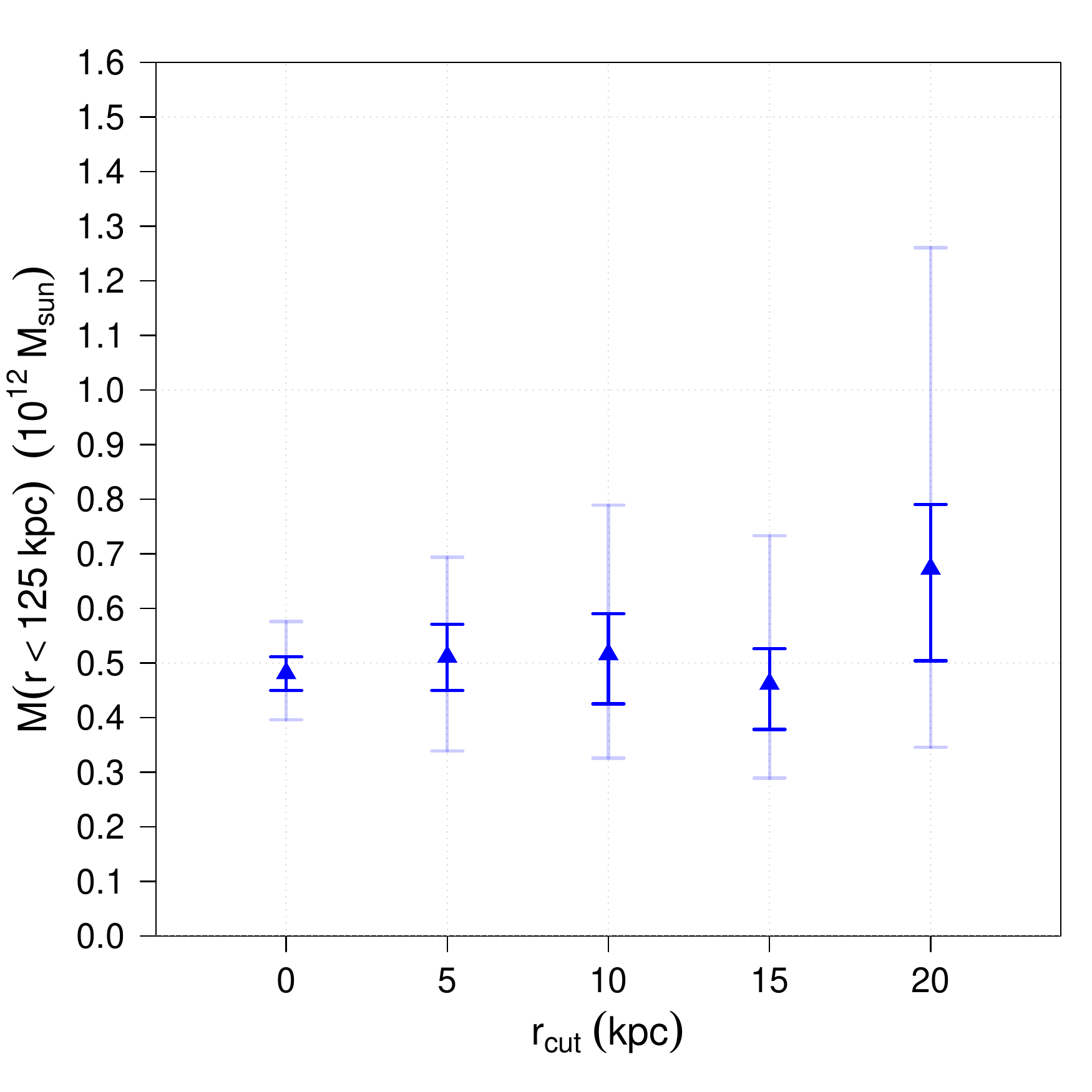}
	\caption{\textbf{Sensitivity analysis:} parameter estimates when GCs within the $r_{cut}$ value are removed from the sample. Inner bars are 50\% credible regions and outer bars are 95\% credible regions.}
	\label{fig:rcut}
\end{figure}
\begin{figure}[h]
\centering  
	\includegraphics[trim={0.5cm 0cm 0cm 0.5cm}, totalheight=0.75\textheight]{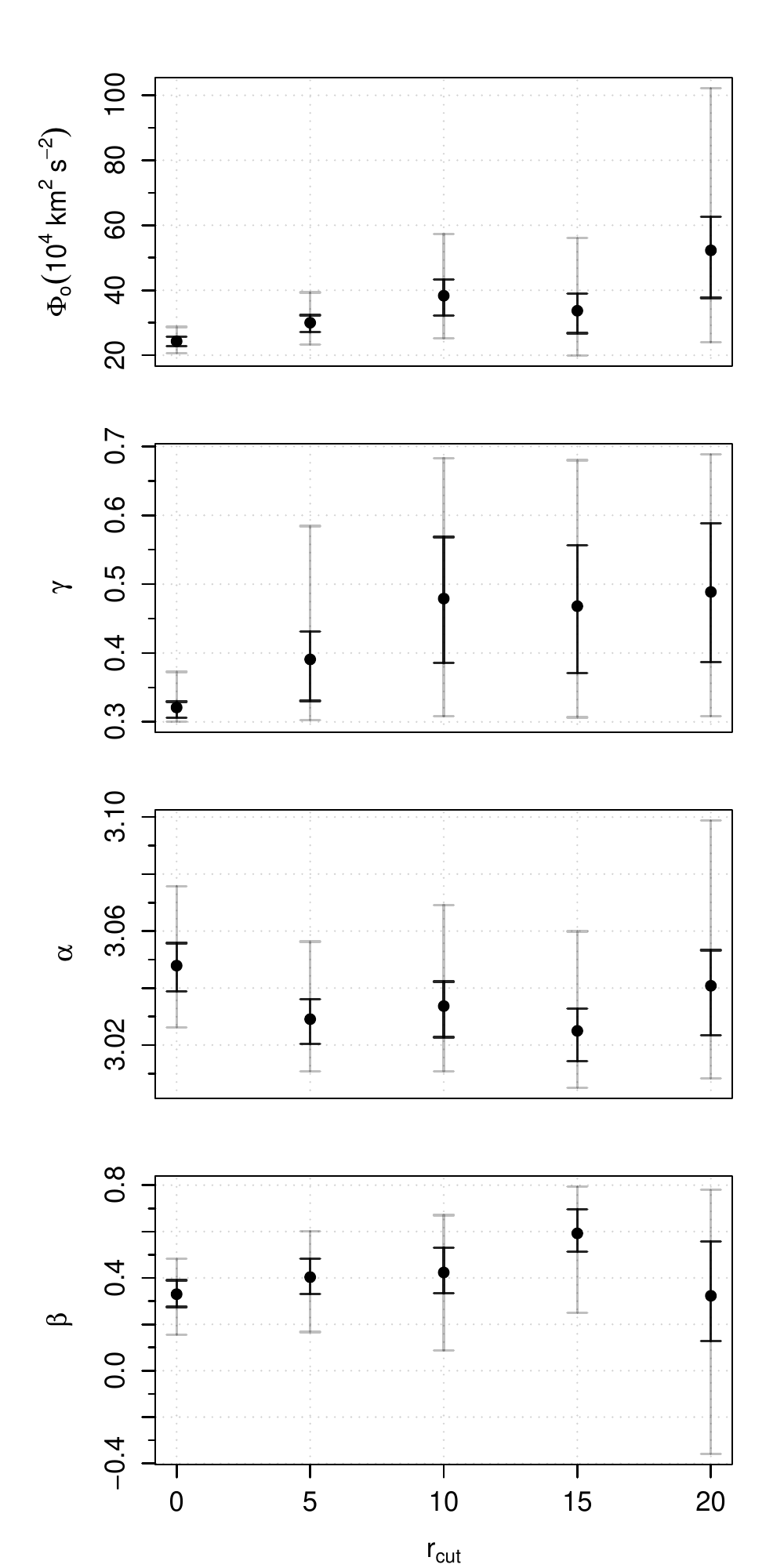}
	\caption{\textbf{Sensitivity analysis:} Parameter estimates when GCs within the $r_{cut}$ value are removed from the sample. Inner bars are 50\% credible regions and outer bars are 95\% credible regions.}
	\label{fig:rcutpars}
\end{figure}

In Paper II, we performed a sensitivity analysis to determine how the spatial sample of GCs might affect the mass estimate of the Milky Way under our assumed power-law model. The sensitivity analysis involved obtaining mass estimates after removing GCs with positions within five different $r_{cut}$ values: 0, 5, 10, 15, and 20 kpc. Here we repeat the sensitivity analysis using the same set of $r_{cut}$ values, but using the full sample of 143 GCs. The sensitivity analysis implicitly examines how disk- and bulge-associated GCs might affect the mass estimate, because when $r_{cut} = 10\kpc$, almost all (93/97) of the excluded GCs have $|z|<5\kpc$.  Figures~\ref{fig:rcut} and \ref{fig:rcutpars} display how the mass and individual model parameters $\Phi_o, \gamma, \alpha, \text{and } \beta$ vary in the sensitivity analysis.

In contrast to Paper II, we find that the mass estimate within 125kpc is robust to the systematic exclusion of inner GCs, except perhaps when only GCs beyond $20\kpc$ are used in the analysis (Figure~\ref{fig:rcut}). We note however that the sample size beyond $20\kpc$ is small (19 GCs), and only 4 of these GCs have proper motion measurements. Accordingly, the uncertainty in the mass increases significantly in this case, and the 95\% credible regions overlap with mass estimates under smaller $r_{cut}$ values.

The results suggest that the current model is adequate for describing the profile of GCs, at least with regard to estimating the total mass within $125\kpc$. If the current tracer model was not able to describe the data, then we might expect changes in the $\alpha$ estimate, and the mass estimate, as inner tracers were systematically removed. However, we see no evidence of this occurring within the 95\% credible regions of the posterior distributions for $\alpha$ (Figure~\ref{fig:rcutpars}), and there is little evidence that changes in $\alpha$ affect the mass estimate (Figure~\ref{fig:rcut}). The power-law slope of the GC population is highly constrained in the analysis, regardless of the GC sample that is used. One thing to note is that the prior $p(\alpha)$ becomes less and less informative for each $r_{cut}$, as the extra data available to define a prior change from 14 GCs to 6, 5, 3, and 3 GCs.

 A positive correlation in the estimates of $\Phi_o$ and $\gamma$ is immediately obvious in the upper two panels of Figure~\ref{fig:rcutpars}, and as more inner GCs are excluded (i.e. as $r_{cut}$ increases) $\gamma \rightarrow 0.5$. This value of $\gamma$ corresponds to an approximate \cite{nfw1996} profile at large radii \citep{deason2011}, albeit with very large uncertainty. The significant change in $\gamma$ and in its uncertainty in the sensitivity analysis suggests
that the shape of the dark matter profile cannot be well constrained using only the outermost GCs. To constrain the shape with more confidence, all of the data must be used. The single power-law profile for the gravitational potential does not take into account the Galaxy's bulge and disk components. However, despite the relatively simplistic model for the gravitational potential and the changes in $\gamma$ during the sensitivity analysis, the mass estimate is robust.


The $\beta$ estimates in the sensitivity analysis are in good agreement with one another, despite the percentage of complete data decreasing as $r_{cut}$ increases.  We can therefore conclude that the GC population has a mildly radial constant anisotropy under this model assumption. However, when the GC sample is limited to clusters outside $20\kpc$ the uncertainty in $\beta$ becomes quite large.

\begin{figure*}
\centering
	\includegraphics[totalheight=0.5\textheight]{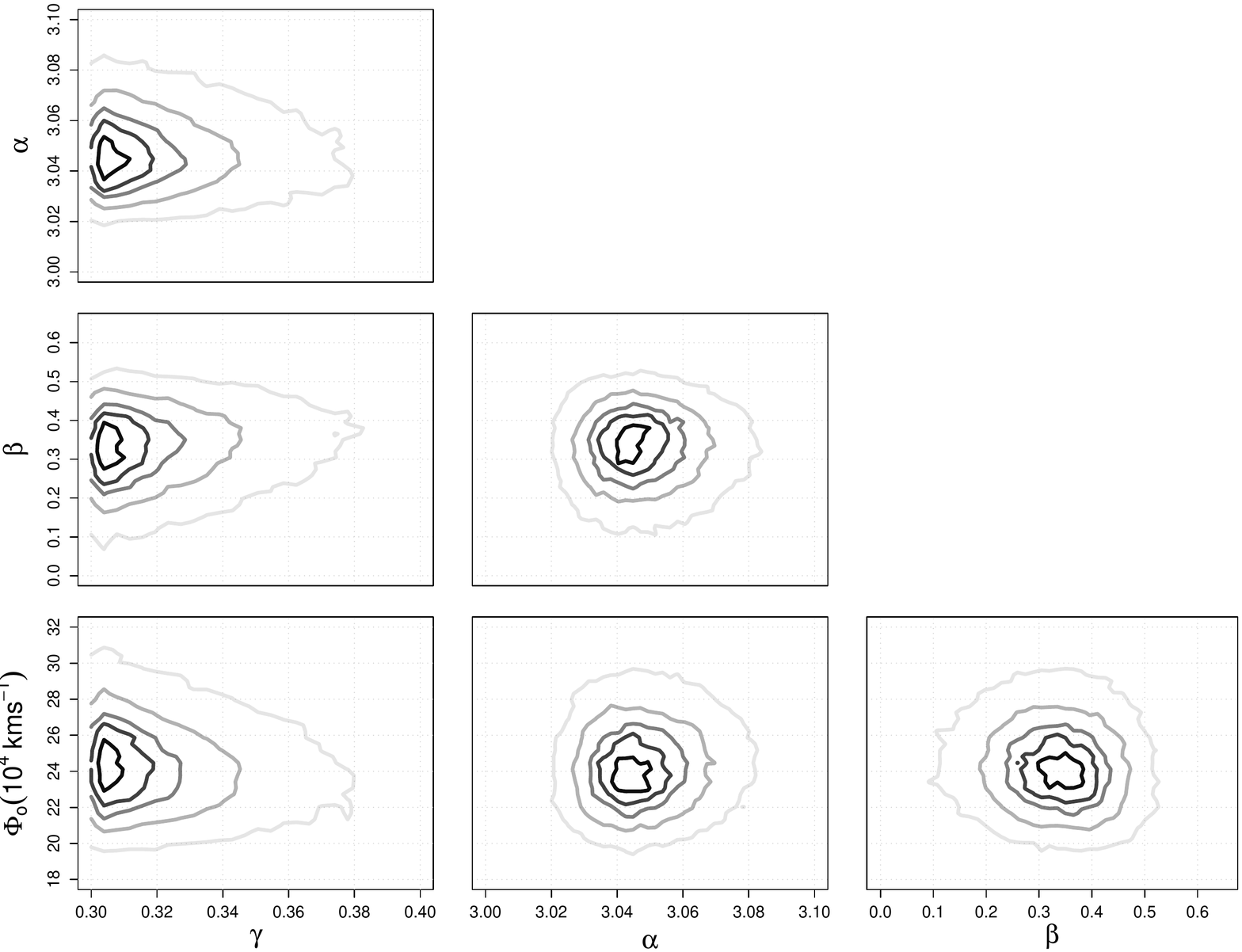}
	\caption{Joint posterior distributions for the model parameters. The curves show the 10, 20, 30, 50, 75, and 95\% Bayesian credible regions.}
	\label{fig:post}
\end{figure*}

To summarize the entire posterior distribution for the full sample of 143 GCs, we also show the joint credible regions for all four model parameters (Figure~\ref{fig:post}).

\begin{figure*}[t]
\centering
\includegraphics[scale=0.7]{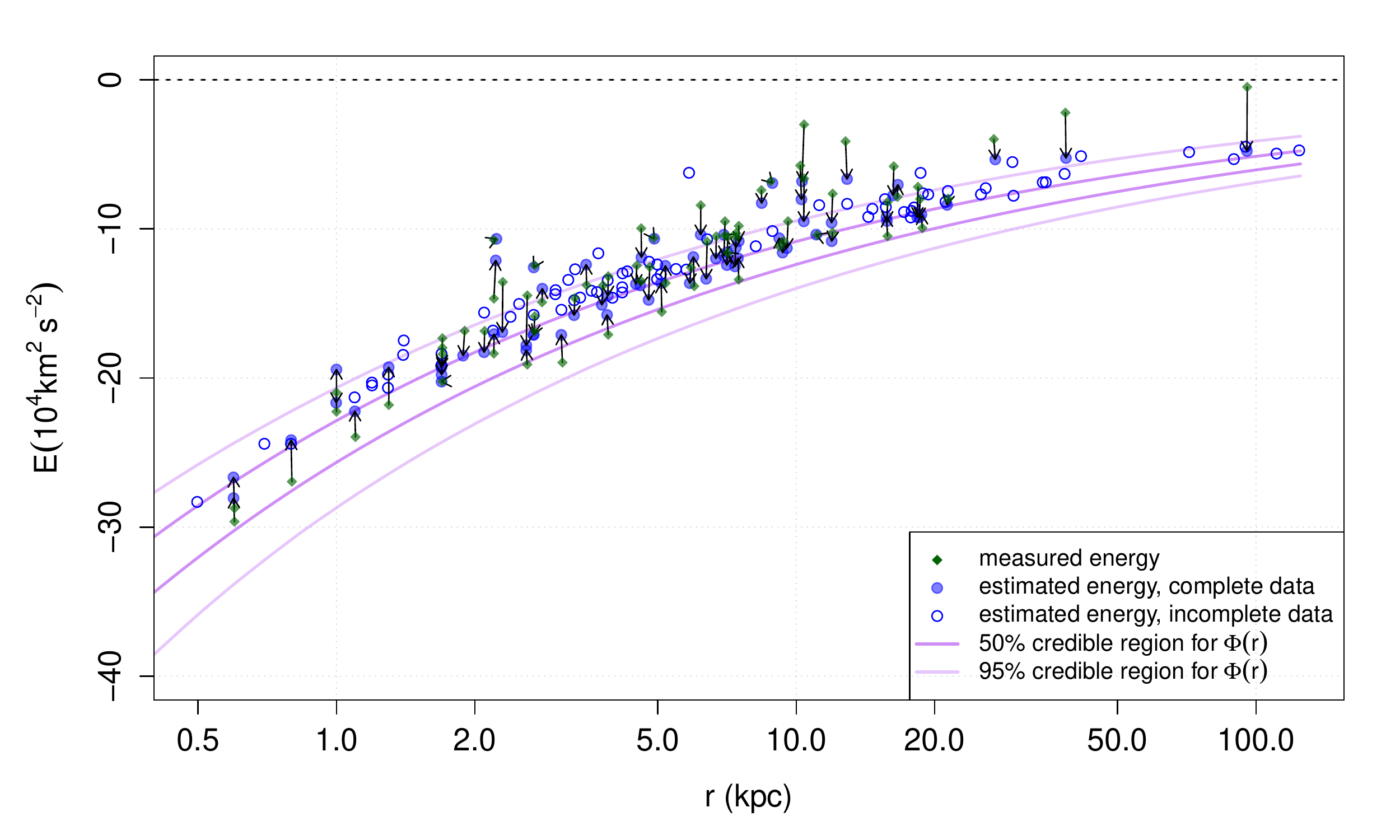}
	\caption{The posterior means of the GCs' specific energies as a function of Galactocentric position. The blue circles are the mean energy estimates for each GC; solid points are complete data and hollow points are incomplete data. The solid green diamonds are the measured energies for complete data. The arrows connect GC measured energies to GC posterior mean energies, and the purple shaded curves represent the 50 and 95\% credible regions for the potential.\vspace{3ex}}\label{fig:energies}
\end{figure*}

Because the hierarchical method treats the Heliocentric distances and velocities as parameters in the model, the final posterior distribution provides estimates and credible regions for the parameters $\bm{\textcolor{blue}{\vartheta}} = \left( \textcolor{blue}{r^{\dagger}},\textcolor{blue}{v^{\dagger}_{los}}, \textcolor{blue} {\mu^{\dagger}_{\delta}}, \textcolor{blue}{\mu_{\alpha}\cos{\delta}^{\dagger}}\right)$ 
 for all 143 GCs (i.e. there are 572 parameters in the GC measurement model alone). Using the posterior distributions for these parameters, we derive an estimate of the specific energy $E$ for each GC, with credible regions. Figure~\ref{fig:energies} shows these energy estimates as a function of Galactocentric position: hollow and solid blue circles are the mean energy estimates of the incomplete and complete data parameters respectively. The solid green diamonds are the energies derived from the measurements of the complete data (there are no hollow green diamonds because energies cannot be derived without proper motions). Arrows from the solid green diamonds to the solid blue points connect the same GC. For legibility, we do not show the 95\% credible regions for the energies, but we have checked that they are reasonable. The shaded purple curves enclose the 50\% and 95\% credible regions for the gravitational potential, determined pointwise as a function of $r$. 

Figure~\ref{fig:energies} provides a consistency check of the hierarchical method in three ways: (1) the distribution of points is consistent with our initial assumptions that all GCs are bound to the Galaxy, (2) the incomplete and complete data energy distributions populate the region between the gravitational potential and the zero line, and (3) the incomplete and complete data do not appear to have different energy distributions. Another feature of note is the tendency for the estimated energies based on positions and velocities to shrink towards a curve similar in shape to the $\Phi(r)$ profile. This is because the posterior distributions for each tracer's energy are in some sense a compromise between the prior implied by the tracer model (Equation~\ref{eq:DFLfinal}) and the measured value. Whether the posterior distribution is closer to the measured value or to the value implied by the tracer model is a function of the width of the prior compared to the measurement uncertainty of the tracer.

\vspace{10pt}
\section{Discussion}

The Bayesian method presented here has an advantage over traditional point mass estimation techniques in the literature because it uses complete and incomplete data simultaneously in the analysis, whereas other techniques use either complete or incomplete data only \citep[e.g. the mass estimators introduced by][]{bahcalltremaine1981, evanswilkinson2003, watkins2010}. Furthermore, although other studies have used a Bayesian analysis to infer the mass of the Milky Way \citep[e.g.][]{little1987, kulessa1992, kochanek1996, wilkinsonevans1999, mcmillan2011, kafle2012, williamsevans2015, kupper2015}, to our knowledge none of these studies has included measurement uncertainties using a coherent measurement model as we have done here.

Including the measurement uncertainties in a measurement model introduced four additional parameters for every GC, which increased the computational cost of the analysis. Nonetheless, even with 576 parameters (572 measurement model parameters $\bm{\textcolor{blue}{\vartheta}}$ and 4 tracer model parameters $\bm{\theta}$), we were able to run these analyses overnight on a personal computer with four cores after sufficient Markov chain burn-in.

We found that including uncertainties in the analysis resulted in a tighter constraint on the cumulative mass profile of the Milky Way compared to ignoring measurement uncertainties (Figure~\ref{fig:89gcs}). This somewhat paradoxical result might be explained by attributing some of the variation in GC kinematics to the measurement process, as described in Figure~\ref{fig:energies}. Without allowing for measurement error, the tracer model is made to explain all of the variation, which apparently results in increased overall uncertainty in the mass profile.

When the sample size of GCs went from 89 to 143, neither the value nor the spread of the mass profile changed substantially (Figure~\ref{fig:143gcs}). Introducing additional data might be expected to decrease the width of the Bayesian credible regions, but this was not observed. We suspect that the credible region width did not change because including additional incomplete data decreased the \emph{proportion} of complete measurements. When 143 GCs were included in the analysis, nearly 50\% of the data were incomplete, in constrast to almost 80\% of the data being complete when 89 GCs were used. We therefore stress the importance of having complete and accurate proper motion data for tracer objects. In particular, there is a need for \emph{remote} tracers with complete measurements. This point is highlighted by both the sensitivity analysis (Figures~\ref{fig:rcut} and \ref{fig:rcutpars}) and by the energy estimates of the GCs (Figure~\ref{fig:energies}). 

Figures~\ref{fig:rcut} and \ref{fig:rcutpars} display how the uncertainty in the mass and parameter estimates changes as inner GCs are removed from the sample; as the percentage of \emph{incomplete} data increases, the results are much less constrained. 

Figure~\ref{fig:energies} shows that the outermost GCs with complete data have estimated energies that are lower than their measurements. The complete data $E$ \emph{estimates} (solid blue circles) appear to move away from the measurement values (solid green diamonds) and towards the $E$ estimates of the \emph{incomplete} data (hollow blue circles). However, there is very little information beyond 20kpc, because the proportion of GCs at large distances without proper motion measurements is high. If complete velocity measurements of these remote GCs suggest that they have high energies, then the mass estimate obtained with this model will increase. If they do not, then the mass estimate of the Galaxy may be closer to the value we found in this study. Ultimately, this question cannot be answered without measuring the proper motions of the other remote GCs.

The results of the sensitivity analysis, the estimated energy profile of the GCs, and the relatively unchanged result between 89 and 143 clusters lead us to conclude that it is absolutely critical to have proper motions for distant tracers. Obtaining proper motions of GCs at large radii is critical to understanding the distribution of energies at large radii and thus the mass of the Milky Way.

An illuminating follow-up investigation to this study is to analyze simulations of Milky Way-type galaxies and their satellites using our hierarchical method. We are currently performing such analyses of realistic galaxy simulations \citep{keller2015, keller2016} to determine how much proper motion data is necessary to constrain the mass profile further, and to study what biases may occur under the Galaxy model that we have employed here when the distribution of the tracers does not follow a single power-law spatial distribution (Eadie, Keller, et al.~in prep).

The mass profile result we have obtained in this study is at the lower end of most mass estimates in the literature, but is also in agreement with some more recent measurements \citep[e.g.][]{deason2012,battaglia2005, gibbons2014}. Because the result obtained in this study is so similar to the mass profile of Paper II, we refer the reader to that paper for further comparison to other studies. We end by noting, however, that our results could change substantially with the inclusion of proper motion data from remote tracers. The number of complete velocity measurements for GCs at large distances will soon increase thanks to projects such as the HST Proper Motion Collaboration (HSTPROMO\footnote{HSTPROMO Project: \url{http://www/stsci.edu/~marel/hstpromo.html}}) \citep{vandermarel2014hstpromo, sohn2016AAS}, and with these data, a better estimate of the Galaxy's total mass will be possible.

\section{Conclusion}

We have described a coherent, hierarchical Bayesian method for estimating the mass profile of the Milky Way Galaxy, and applied this method to the Galaxy using GC data. This statistical framework allows us to take full advantage of all of the available GC kinematic data, and also provides a meaningful and coherent probabilistic way to incorporate measurement uncertainties. 

Using the assumptions of the power-law model (Section~\ref{sec:df}), the hierarchical framework for including uncertainties (Section~\ref{sec:methods}), and the prior distributions (Section~\ref{sec:priors}), and confronting this model with data from 143 GCs around the Milky Way, we arrive at a cumulative mass profile for the Galaxy with uncertainties (Figure~\ref{fig:143gcs}) and a mass estimate within $125\kpc$ of $4.8\times10^{11}\msun$ (the 95\% Bayesian credible regions are $(4.0-5.8)\times10^{11}\msun$). When we extrapolate the mass profile to the virial radius $(\approx 179\kpc)$, we find $M_{vir}=6.2\times10^{11}\msun$ with a 95\% Bayesian credible region of $(4.7-7.8)\times10^{11}\msun$. This mass estimate is notably lower than those in other studies.

The statistical framework presented here will be highly useful and appropriate for other tracer objects around the Milky Way, such as halo stars and DGs. Using our approach with data sets from large programs, such as Gaia, could yield a well-constrained mass estimate for the Galaxy. Incorporating large data sets in this analysis will present some computational challenges, but given the effectiveness of our MCMC sampler we are confident that this will be a tractable problem through parallelization. 

The first order of business, however, is to better understand what tracer populations will provide the most information about the Milky Way's gravitational potential. Thus, in our next paper Eadie, Keller, et al. (in preparation) we perform a series of blind tests of simulated data of Milky Way-type galaxies that were created through hydrodynamical simulations \citep{keller2015,keller2016}, and investigate which tracer information is necessary for constraining the mass of the Milky Way.

\acknowledgements

WEH and GME acknowledge the financial support of McMaster University and NSERC. The authors would also like to thank the referee for thoughtful comments and suggestions.

\bibliographystyle{aasjournal}

\bibliography{myrefsnew}

\label{lastpage}

\end{document}